%% file: main.tex
\documentclass{article}

\usepackage{spconf,amsmath,amssymb,graphicx}
\usepackage[hidelinks]{hyperref}
\usepackage{booktabs,multirow,xcolor,placeins,afterpage,float}

\makeatletter
\let\leantts@thebibliography\thebibliography
\renewcommand{\thebibliography}[1]{%
  \leantts@thebibliography{#1}%
  \fontsize{9}{9.2}\selectfont
  \setlength{\itemsep}{0pt}%
  \setlength{\parsep}{0pt}%
  \setlength{\topsep}{0pt}%
  \setlength{\partopsep}{0pt}%
}
\makeatother

\title{LeanTTS: Efficient Full-Stack Bilingual TTS for\\
Zero-Shot Cloning and Voice Customization}

\name{\begin{tabular}{c}
Yuheng Lu$^{1,*}$, Qiulin Li$^{2,*}$, Xi Wang$^{3,*}$, Yang Wang$^{2}$\\[-1pt]
Shengpei Jiang$^{2}$, Liyin Liang$^{2}$, Yifeng Qian$^{2}$, Longbiao Wang$^{1}$
\end{tabular}\vspace{-1.0em}
\thanks{$^{*}$Equal contribution.}}

\address{
{\small $^{1}$Tianjin University, Tianjin, China \quad
$^{2}$Didichuxing Co. Ltd., Beijing, China \quad
$^{3}$Tsinghua University, Beijing, China}\\[4pt]
{\small\texttt{\{luyuheng2024,wanglongbiao\}@tju.edu.cn}}\\
{\small\texttt{\{qiunnlili,ianwangyang,jiangshengpei,liangliyinkevin,qianyifeng\}@didiglobal.com}}\\
{\small\texttt{xi-wang24@mails.tsinghua.edu.cn}}}

\begin{document}
\ninept
\maketitle
\begin{abstract}
While recent advances in text-to-speech (TTS) synthesis have substantially improved speech quality and modeling flexibility, the development of lightweight and computationally efficient systems remains comparatively underexplored. We present LeanTTS, a bilingual complete synthesis
stack that supports two complementary regimes: reference-prompt zero-shot
cloning for unseen speakers and token-conditioned customization for recurring
voices. LeanTTS combines a compact Mandarin polyphone front end, a structurally
compressed flow-matching acoustic model, and a lightweight Vocos decoder with
four-step MeanFlow generation. The resulting system contains 18.9M learned
parameters, 86.12\% fewer than the 136.2M-parameter ZipVoice--Vocos pipeline.
On Seed-TTS-Eval, its zero-shot mode obtains 2.02 CER/62.3 SIM on Mandarin and
2.32 WER/48.3 SIM on English, exposing the quality--size trade-off under
aggressive compression. With token-conditioned customized voices, the same
system reaches 1.57 CER/77.2 SIM on Mandarin and 1.93 WER/76.6 SIM on English,
with UTMOS scores of 3.03 and 4.33, respectively. It further achieves
mean/P90 CPU RTFs of 0.049/0.061, establishing a sub-20M operating point for
bilingual personalization and real-time CPU deployment. Audio demos are
available at \url{https://github.com/notlu/leantts}.
\end{abstract}


\begin{figure*}[t]
  \centering
  \includegraphics[
    width=\textwidth
  ]{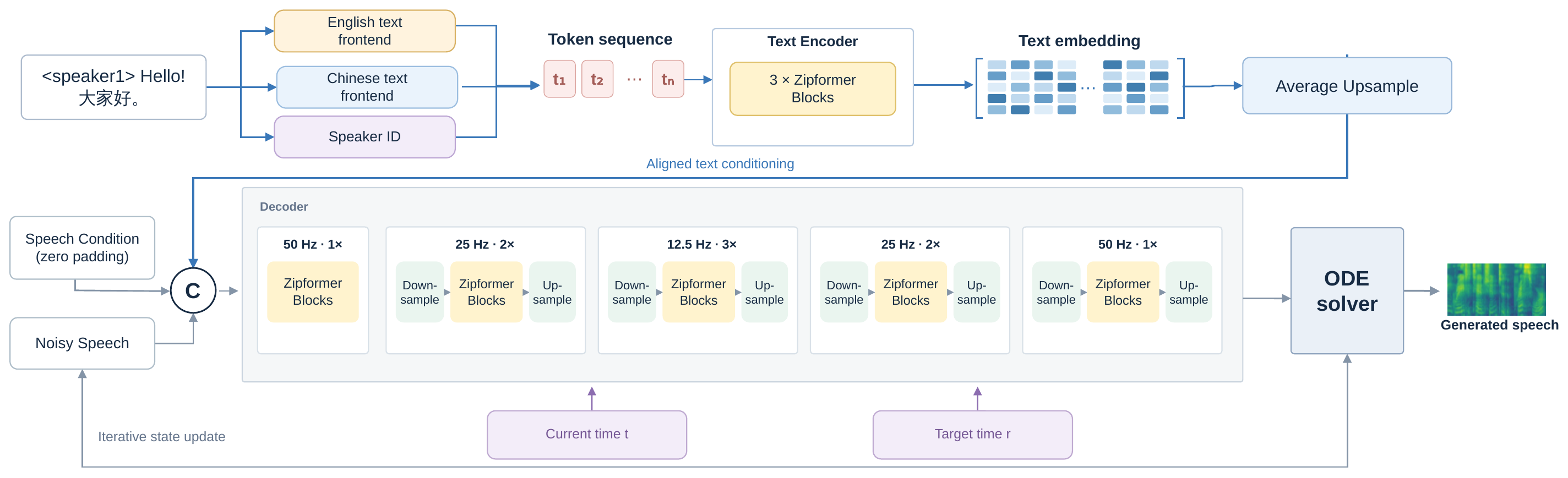}
  \vspace{-14pt}
  \caption{Bilingual preset-voice inference and acoustic-model architecture of
  LeanTTS.}
  \label{fig:pipeline}
\end{figure*}

\section{Introduction}
\label{sec:introduction}

Recent advances in generative text-to-speech (TTS) have enabled high-quality
synthesis with increasingly flexible capabilities, including multilingual
generation and zero-shot adaptation to unseen speakers~\cite{zhu2025zipvoice,
du2024cosyvoice2,du2025cosyvoice3,chen2024f5tts}. However, these advances are
often accompanied by rapidly growing model size and computational cost. Modern
zero-shot TTS systems commonly rely on hundreds of millions or even billions
of parameters, making them difficult to deploy on resource-constrained devices
where memory, storage, and latency are tightly limited. This is particularly
undesirable for personalized applications such as in-vehicle assistants and
embedded devices, where local synthesis can reduce network dependency and
avoid transmitting speaker-specific reference audio to remote servers.

A substantial body of work has therefore explored lightweight and efficient
TTS through compact architectures, neural architecture search, streaming
synthesis, and efficient waveform generation~\cite{popov2020ondevice,
luo2021lightspeech,shopov2023streamspeech,atienza2023efficientspeech,
guo2024flytts,fujita2024mixture,chen2025lightstable,lee2025smam,
kim2025supertonic,lemerle2024smalle}. Nevertheless, most existing compact
systems focus on fixed-speaker, single-voice, or otherwise constrained
synthesis settings. In contrast, recent zero-shot TTS systems provide much
greater flexibility in speaker adaptation, but typically retain a considerably
larger computational footprint~\cite{zhu2025zipvoice,du2024cosyvoice2,
du2025cosyvoice3,chen2024f5tts}. This leaves an important gap between
\emph{efficient on-device synthesis} and \emph{modern personalized TTS}: a
practical edge system should remain compact enough for real-time local
inference while supporting both unseen-speaker synthesis and high-quality
customized voices. Moreover, deployment cost is determined by the complete
synthesis pipeline rather than the acoustic model alone, including
pronunciation modeling, acoustic generation, and waveform reconstruction.

To bridge this gap, we present \textbf{LeanTTS}, a compact full-stack bilingual
TTS system designed for efficient personalized synthesis. LeanTTS jointly
reduces the Mandarin pronunciation front end~\cite{chen2022g2pw}, the
flow-matching acoustic generator, and the neural waveform decoder
~\cite{siuzdak2023vocos}, instead of optimizing these components in isolation.
The system supports two complementary usage modes: reference-prompt
\emph{zero-shot voice cloning} for previously unseen speakers, and
token-conditioned \emph{preset-voice customization} for frequently used target
voices. We further introduce a compact acoustic training pipeline with
teacher-guided recovery and MeanFlow continuation~\cite{geng2025meanflow},
enabling few-step generation while maintaining the flexibility of the original
flow-matching formulation.

Our deployment-focused LeanTTS configuration contains only 18.9M learned
parameters, reducing the complete ZipVoice--Vocos synthesis path
~\cite{zhu2025zipvoice,siuzdak2023vocos} by 86.12\% while retaining
Mandarin--English zero-shot synthesis. For customized preset voices, the
four-step model achieves 1.57 CER/77.2 SIM on Mandarin and 1.93 WER/76.6 SIM
on English. On CPU, the same compact system reaches a mean/P90 real-time factor
of 0.049/0.061 at four synthesis steps. These results demonstrate that a
complete personalized TTS stack can be aggressively compressed while still
providing a practical operating point for bilingual zero-shot cloning, voice
customization, and real-time local deployment.

\section{Method}
\label{sec:method}

\subsection{Full-stack overview}

Figure~\ref{fig:pipeline} illustrates the bilingual preset-voice inference path
and the internal architecture of the text encoder and FM decoder. A leading
speaker tag selects the preset voice, while language-specific front ends convert
Mandarin and English spans into pronunciation tokens. The compact acoustic
generator predicts a Mel spectrogram, and Vocos reconstructs the 24-kHz
waveform. We build a 25M bilingual flow-matching model and a smaller 13M FM
model, then obtain a separate few-step checkpoint through MeanFlow
continuation.

\textbf{Preset-speaker conditioning.}
We append eight speaker tags at IDs 360--367 to the original 360-entry
phoneme-and-punctuation vocabulary. A speaker tag occupies the first input
position, bypasses G2P, and
is jointly encoded with the pronunciation tokens into the text condition.
Training progressively removes the reference-speech acoustic condition for
tagged examples while retaining the text condition, enabling token-only preset
voice inference without affecting reference-prompt zero-shot cloning.

\subsection{Compact Mandarin pronunciation front-end}

For Mandarin, contextual modeling is invoked only for ambiguous polyphones.
Deterministic G2P provides candidate pronunciations, while characters outside
the target polyphone inventory directly retain the rule-based prediction.

For each ambiguous position, a two-layer contextual encoder processes a
target-centered window of at most 42 characters and produces a
128-dimensional representation.\footnote{\url{https://huggingface.co/uer/roberta-tiny-wwm-chinese-cluecorpussmall}}
The prediction module covers 552 polyphonic characters and 659 global
pronunciation labels. It retains a shared $659\times128$ pronunciation bank,
but does not materialize all 659 logits. Instead, a fixed $552\times6$
candidate-index buffer gathers at most six pronunciation rows for the target
character, and a mask removes unused slots. Six-dimensional global-slot,
character, and character--part-of-speech terms condition the resulting local
logits, while a part-of-speech classifier provides auxiliary syntactic
supervision. The resulting prediction is followed by deterministic
tone-sandhi processing and converted to acoustic-model tokens. The complete
candidate-slot G2PW contains 3.3M learned parameters.

\subsection{Structured acoustic compression and recovery}

We derive two compact acoustic models from the 122.7M-parameter teacher using
stage-wise depth reduction and L2-norm channel selection. The teacher contains
five FM-decoder stages with $[2,2,4,4,4]$ Zipformer blocks, a decoder model/FFN width of $512/1536$, and a four-layer text encoder
with a model/FFN width of $192/512$. The 25M configuration contains 25.1M
parameters: its decoder is reduced to $[1,2,3,2,1]$ blocks with
a model/FFN width of $384/384$, while its text encoder uses three layers with
a model/FFN width of $192/384$. The 13M configuration contains 13.3M
parameters and retains the same $[1,2,3,2,1]$ decoder depths and three-layer
text encoder, but further narrows the decoder model/FFN width to $256/256$;
the text-encoder model/FFN width remains $192/384$. Both students preserve
the teacher's stage-wise downsampling factors.

For each reducible hidden channel $j$, its importance is measured by
\begin{equation}
 s_j=\lVert W_{\mathrm{in}}[j,:]\rVert_2^2+
     \lVert W_{\mathrm{out}}[:,j]\rVert_2^2+b_j^2.
 \label{eq:l2_score}
\end{equation}
The highest-scoring channels are retained consistently across coupled
projections, while compatible parameters are inherited directly from the
teacher. The resulting students remain regular dense networks.

After structural transfer, the student is progressively recovered using
teacher guidance and a ground-truth CFM objective:
\begin{equation}
 \mathcal{L}_{\mathrm{recover}}
 =
 w_{\mathrm{KD}}\mathcal{L}_{\mathrm{vKD}}
 +
 w_{\mathrm{GT}}\mathcal{L}_{\mathrm{CFM}} .
 \label{eq:recover}
\end{equation}
During recovery, $w_{\mathrm{KD}}$ decreases from 1.0 to 0.2 while
$w_{\mathrm{GT}}$ increases from 0 to 1.0. Training then continues with
ground-truth flow matching alone, allowing the compressed initialization to
first recover teacher behavior and subsequently adapt to the data target.

\subsection{Fixed-grid rollout distillation}

We post-train the recovered 13M model with a frozen flow-matching teacher on
the fixed four-step inference grid from $t=1$ (noise) to $t=0$ (data), with
step size 0.25. For each adjacent interval $(t,r)$, the student predicts an
average velocity and applies
\begin{equation}
 z_r=z_t-(t-r)u_\theta(z_t,t,r,c).
 \label{eq:meanflow}
\end{equation}

Thus, one network evaluation replaces the teacher trajectory over the
corresponding interval rather than approximating an instantaneous
velocity.

Training samples only these four intervals. To expose the student to
accumulated inference errors, we mix prompt-preserving clean states with
stop-gradient states obtained by rolling out its preceding grid steps; the
on-policy probability is gradually increased to 0.5. From the resulting
state $z_t$, the teacher performs an eight-step FP32 Euler rollout to $y_r^T$,
yielding the state-conditioned target $u^T=(z_t-y_r^T)/(t-r)$. We minimize
masked MSE between $u_\theta$ and $u^T$ only on valid generated frames, while
prompt and padding frames remain fixed.

\input{table_main_results}

\subsection{Compact Vocos}

We compress Vocos while preserving its 100-bin Mel and 24-kHz waveform
interfaces. The eight-block architecture is retained, while the backbone
width is reduced from 512 to 192 and the intermediate width from 1536 to
576, yielding a 2.1M-parameter decoder. The compact model is retrained using
adversarial, feature-matching, and multi-resolution spectral objectives.

\section{Experiments}
\label{sec:experiments}

\subsection{Experimental setup}

\textbf{Data.}
The acoustic models are trained on Emilia~\cite{he2024emilia}.
Both acoustic variants use Mandarin and English data. The 13M checkpoint
jointly learns eight preset-speaker tokens and is evaluated under both prompt
and token conditioning. Each preset speaker is represented by approximately
300 hours of Mandarin and 300 hours of English training speech. The compact
Vocos is trained on approximately 2,245 hours of 24-kHz audio. The
pronunciation front-end follows the CVTE-Poly benchmark~\cite{zhang2023cvtepoly}:
training uses CPP and CVTE-Poly, and evaluation uses the refined CPP test split.

\textbf{Evaluation.}
We evaluate zero-shot synthesis on Seed-TTS-Eval
~\cite{anastassiou2024seedtts} using three reproducible model-based metrics.
For component evaluation, we follow CVTE-Poly~\cite{zhang2023cvtepoly} and
report overall accuracy (Acc.), character-macro accuracy (Acc.avg.p), and
character--pronunciation-macro accuracy (Acc.avg.pp) for the pronunciation
front end. Following Vocos~\cite{siuzdak2023vocos}, vocoder reconstruction is
evaluated using UTMOS, ViSQOL, PESQ, voiced/unvoiced F1, and periodicity error.
For intelligibility, we compute the error rate between the input text and an
ASR transcription of the synthesized speech: Whisper large-v3
~\cite{radford2023whisper}\footnote{\url{https://github.com/openai/whisper}}
is used for test-en, and SeACo-Paraformer Large
~\cite{shi2023seaco}\footnote{\url{https://github.com/R1ckShi/SeACo-Paraformer}}
for test-zh, yielding WER and CER, respectively. Speaker similarity (SIM) is
the cosine similarity between embeddings of the prompt and synthesized speech,
extracted by a WavLM-Large front end~\cite{chen2022wavlm} with an ECAPA-TDNN
speaker model~\cite{desplanques2020ecapa}\footnote{\url{https://github.com/microsoft/UniSpeech/tree/main/downstreams/speaker_verification}}.
Naturalness is estimated with the UTMOS22 strong learner
~\cite{saeki2022utmos}\footnote{\url{https://github.com/sarulab-speech/UTMOS22}}.
Efficiency is reported using learned parameters and real-time factor (RTF),
where lower RTF indicates faster synthesis. We report both the mean and the
90th-percentile (P90) RTF. To assess compatibility with efficient on-device
speech systems, we additionally export the four-step LeanTTS-13M pipeline to
the MNN mobile inference framework\footnote{\url{https://github.com/alibaba/MNN}}
and integrate it into an Android application executed in an Android emulator.

\textbf{Baselines.}
Among publicly documented systems with complete deployment-size accounting,
we did not identify a sub-100M baseline that jointly reports bilingual
zero-shot voice cloning and directly comparable evaluation. Compact TTS
systems below this scale are predominantly monolingual and fixed- or
multi-speaker, whereas comparable zero-shot systems are substantially larger.
We therefore compare zero-shot synthesis against ZipVoice, CosyVoice3,
Spark-TTS, and MOSS-TTS-Nano, and compare preset/fixed-voice synthesis against
the compact Inflect-Micro-v2 and Matcha-TTS systems. Component studies further
compare the original G2PW and Vocos models with their compact counterparts.

\subsection{Overall system comparison}

Table~\ref{tab:main_results} jointly compares complete learned-parameter counts
and synthesis quality. LeanTTS-13M-FM is evaluated at 16 steps, whereas the
separately continued LeanTTS-13M-MF checkpoint is evaluated at four steps.
Table~\ref{tab:main_results} separates reference-prompt zero-shot cloning from
preset/fixed-voice synthesis so that systems are compared under the same
conditioning setting. Preset-token metrics are averaged over the eight learned
speakers. LeanTTS-25M retains conventional 16-step flow-matching inference for
the quality evaluation and serves as a higher-capacity bilingual reference.
Four-NFE CPU latency is reported separately in Table~\ref{tab:rtf}.

\begin{table}[!t]
  \caption{Component-wise CPU real-time factor (RTF) at four NFEs. Mean and
  90th-percentile (P90) RTF are reported; lower is faster.}
  \label{tab:rtf}
  \vspace{3pt}
  \centering
  \scriptsize
  \setlength{\tabcolsep}{3pt}
  \resizebox{\columnwidth}{!}{%
  \begin{tabular}{llccc}
    \toprule
    Acoustic model & Vocoder & NFE & Mean RTF $\downarrow$ & P90 RTF $\downarrow$ \\
    \midrule
    ZipVoice-distill & Original Vocos & 4 & 1.6975 & 2.2617 \\
    ZipVoice-distill & Compact Vocos & 4 & 1.6190 & 2.1540 \\
    LeanTTS-25M & Compact Vocos & 4 & 0.0610 & 0.0720 \\
    LeanTTS-13M & Compact Vocos & 4 & 0.0490 & 0.0610 \\
    \bottomrule
  \end{tabular}}
  \\[2pt]
  \parbox{0.96\columnwidth}{\tiny
  Protocol: all 2,020 Seed-TTS-Eval test-zh
  utterances; MacBook Pro, Apple M5 (10 cores), 16 GB, macOS 26.5 arm64;
  CPU-only, no GPU/MPS, one PyTorch thread; Python 3.12.13, PyTorch/torchaudio
  2.8.0, NumPy 2.5.3, qnnpack; compact Vocos INT8 checkpoint dequantized on
  loading.}
\end{table}

Table~\ref{tab:rtf} separates the latency gains from vocoder and acoustic-model
compression. Replacing the original Vocos with Compact Vocos reduces the
mean/P90 RTF of ZipVoice-distill from 1.6975/2.2617 to 1.6190/2.1540, showing a
modest vocoder-side improvement. The substantially larger reductions to
0.0610/0.0720 and 0.0490/0.0610 for LeanTTS-25M and LeanTTS-13M, respectively,
indicate that the dominant RTF gain comes from reducing the acoustic-model
size.

The MNN-based Android implementation for LeanTTS-25M achieves an end-to-end RTF of 0.0327 in
the emulator. 

LeanTTS-25M contains a 3.3M front end, a 25.1M acoustic model, and a 2.1M
vocoder, totaling 30.6M learned parameters; LeanTTS-13M contains 18.9M.

\textbf{Zero-shot regime.} At four NFEs, LeanTTS-13M obtains 2.02 CER/62.3 SIM
on Mandarin and 2.32 WER/48.3 SIM on English. Relative to the 136.2M ZipVoice
pipeline, the lower similarity makes the cost of aggressive compression
explicit, while the 86.12\% parameter reduction and 0.049 mean CPU RTF define
a practical quality--size--latency operating point.

\textbf{Customized-voice regime.} For voices represented during training, the
same four-step model reaches 1.57 CER/77.2 SIM/3.03 UTMOS on Mandarin and
1.93 WER/76.6 SIM/4.33 UTMOS on English. Token conditioning therefore raises
SIM by 14.9 and 28.3 points over reference-prompt zero-shot synthesis in the
two languages, respectively, while requiring no reference recording at
inference. LeanTTS-25M provides a higher-capacity zero-shot point, obtaining
1.52/1.75 CER/WER, 65.2/52.3 SIM, and 2.62/3.32 UTMOS on
Mandarin/English.

\subsection{Component analysis}

\textbf{Mandarin front-end.}
Table~\ref{tab:frontend_comparison} compares candidate-slot G2PW with compact
and 8L/512 global-head counterparts under the same evaluation protocol.

Reducing the contextual encoder from 8L/512 to 2L/128 causes a moderate
accuracy decrease, but the compact models remain within an acceptable range.
More importantly, Candidate-slot G2PW performs similarly to the compact
global-head counterpart (98.0631\% versus 98.1527\% overall accuracy) while
substantially reducing the pronunciation-head and character-conditioned
parameters by scoring at most six valid candidates instead of the full global
pronunciation inventory.

\begin{table}[!t]
  \caption{Mandarin polyphone disambiguation under a common evaluation
  protocol.}
  \label{tab:frontend_comparison}
  \centering
  \scriptsize
  \setlength{\tabcolsep}{2.5pt}
  \resizebox{\columnwidth}{!}{%
  \begin{tabular}{lrrr}
    \toprule
    Front end & Overall $\uparrow$ & Acc.avg.p $\uparrow$ & Acc.avg.pp $\uparrow$ \\
    \midrule
    Candidate-slot G2PW & 98.0631 & 97.5011 & 94.2679 \\
    Compact global-head G2PW & 98.1527 & 97.5377 & 94.5919 \\
    Global-head G2PW (8L/512) & 98.7125 & 98.3513 & 96.4304 \\
    \bottomrule
  \end{tabular}%
  }
\end{table}

\textbf{Vocoder.}
Table~\ref{tab:vocoder_ablation} evaluates waveform reconstruction independently
of the acoustic model. Reducing Vocos from 13.5M to 2M parameters causes a
measurable degradation: for the FP32 model, UTMOS decreases from 3.7466 to
3.5116, ViSQOL from 4.6584 to 4.5365, and PESQ from 3.6675 to 3.3723. INT8
quantization further lowers these scores, but substantially reduces the
deployment footprint of the 2M model while retaining acceptable reconstruction
quality. The results therefore expose an explicit trade-off between waveform
quality and deployment cost.

\begin{table}[!t]
  \caption{Vocoder reconstruction on LibriTTS test-clean.}
  \label{tab:vocoder_ablation}
  \vspace{3pt}
  \centering
  \scriptsize
  \setlength{\tabcolsep}{3pt}
  \resizebox{\columnwidth}{!}{%
  \begin{tabular}{lrrrrr}
    \toprule
    Vocos & UTMOS $\uparrow$ & ViSQOL $\uparrow$ & PESQ $\uparrow$ & V/UV F1 $\uparrow$ & Period. $\downarrow$ \\
    \midrule
    13.5M FP32 & 3.7466 & 4.6584 & 3.6675 & 0.9585 & 0.1007 \\
    2M FP32 & 3.5116 & 4.5365 & 3.3723 & 0.9497 & 0.1203 \\
    2M INT8 & 3.1921 & 4.3623 & 3.1963 & 0.9405 & 0.1339 \\
    \bottomrule
  \end{tabular}%
  }
\end{table}

\subsection{Impact of Teacher-Guided Recovery}

We isolate the effect of teacher guidance using the same compressed
architecture, training data, and number of inference steps. As shown in
Table~\ref{tab:teacher_guidance_ablation}, teacher-guided recovery and
ground-truth FM continuation are comparable during the early recovery stage.
The difference emerges with subsequent training: the teacher-guided trajectory
reaches stronger later checkpoints, including simultaneous error-rate and SIM gains
in both languages at 300k and 500k updates. This suggests that teacher
guidance primarily provides a better recovery initialization rather than an
immediate advantage, allowing the compressed model to realize greater
performance potential during continued optimization.

\begin{table}[!t]
  \caption{Recovery-objective ablation at matched training updates and
  inference steps. Each entry reports ground-truth FM only
  $\rightarrow$ teacher-guided recovery; bold indicates the better result.}
  \label{tab:teacher_guidance_ablation}
  \vspace{3pt}
  \centering
  \scriptsize
  \setlength{\tabcolsep}{2.5pt}
  \resizebox{\columnwidth}{!}{%
  \begin{tabular}{lcccc}
    \toprule
    Update & EN WER $\downarrow$ & EN SIM (\%) $\uparrow$ &
    ZH CER $\downarrow$ & ZH SIM (\%) $\uparrow$ \\
    \midrule
    250k & $\mathbf{2.067}\!\rightarrow\!2.108$ &
    $49.6\!\rightarrow\!\mathbf{50.2}$ &
    $\mathbf{1.658}\!\rightarrow\!1.730$ &
    $63.5\!\rightarrow\!\mathbf{63.8}$ \\
    300k & $2.256\!\rightarrow\!\mathbf{1.937}$ &
    $48.4\!\rightarrow\!\mathbf{50.5}$ &
    $1.933\!\rightarrow\!\mathbf{1.646}$ &
    $62.2\!\rightarrow\!\mathbf{64.4}$ \\
    450k & $1.893\!\rightarrow\!\mathbf{1.782}$ &
    $51.2\!\rightarrow\!\mathbf{51.6}$ &
    $1.664\!\rightarrow\!\mathbf{1.623}$ &
    $\mathbf{64.7}\!\rightarrow\!63.6$ \\
    500k & $1.975\!\rightarrow\!\mathbf{1.742}$ &
    $50.7\!\rightarrow\!\mathbf{52.1}$ &
    $1.701\!\rightarrow\!\mathbf{1.584}$ &
    $63.2\!\rightarrow\!\mathbf{64.7}$ \\
    \bottomrule
  \end{tabular}}
\end{table}

\FloatBarrier
\section{Conclusion}
\label{sec:conclusion}

LeanTTS demonstrates that a sub-20M complete bilingual TTS stack can serve two
complementary personalization regimes. Reference-prompt inference retains
zero-shot cloning for unseen speakers, with the expected quality trade-off of
aggressive compression: at four NFEs, the 18.9M system obtains 2.02 CER/62.3
SIM on Mandarin and 2.32 WER/48.3 SIM on English. When recurring voices are
available during training, token-conditioned customization raises similarity
to 77.2 and 76.6 while obtaining 1.57 Mandarin CER and 1.93 English WER. The
same four-step system achieves mean/P90 CPU RTFs of 0.049/0.061. These results
establish a practical operating point across speaker coverage, voice stability,
model size, and latency. Current evaluation is limited to an Apple M5 CPU and
eight customized voices. An MNN-based Android-emulator deployment further
achieves an end-to-end RTF of 0.0327; validation on physical mobile SoCs and
broader multilingual voice sets remains future work.

\clearpage
\bibliographystyle{IEEEbib}
\bibliography{strings,refs}

\end{document}

%% file: table_main_results.tex
\begin{table*}[!t]
  \caption{System-level comparison on Seed-TTS-Eval. Mandarin intelligibility
  is reported as CER and English intelligibility as WER; baseline values follow
  the corresponding official reports.}
  \label{tab:main_results}
  \centering
  \begingroup
  \fontsize{9}{9.5}\selectfont
  \setlength{\tabcolsep}{3pt}
  \renewcommand{\arraystretch}{0.94}
  \begin{tabular*}{\textwidth}{@{\extracolsep{\fill}}lcccccccc@{}}
    \toprule
    & & & \multicolumn{6}{c}{Objective metrics} \\
    \cmidrule(lr){4-9}
    & & & \multicolumn{3}{c}{test-zh} &
    \multicolumn{3}{c}{test-en} \\
    \cmidrule(lr){4-6}\cmidrule(lr){7-9}
    Model & NFE & Params &
    CER $\downarrow$ & SIM $\uparrow$ & UTMOS $\uparrow$ &
    WER $\downarrow$ & SIM $\uparrow$ & UTMOS $\uparrow$ \\
    \midrule
    \multicolumn{9}{@{}l}{\textit{Zero-shot voice cloning}} \\
    CosyVoice3~\cite{du2025cosyvoice3} & 10 &
    $\geq$1.8B & 1.12 & 78.1 & 3.31 & 2.22 & 72.0 & 3.97 \\
    Spark-TTS~\cite{wang2025spark} & -- &
    662.989M & 1.21 & 67.1 & 3.27 & 1.97 & 58.7 & 3.94 \\
    MOSS-TTS-Nano~\cite{gong2026mosstts} & -- &
    139.281M & 4.40 & 64.1 & 2.96 & 3.09 & 49.9 & 3.62 \\
    ZipVoice~\cite{zhu2025zipvoice} & 16 & 136.196M &
    1.42 & 75.2 & 3.16 & 1.67 & 69.6 & 3.83 \\
    LeanTTS-25M & 16 & 30.557M &
    1.52 & 65.2 & 2.62 & 1.75 & 52.3 & 3.32 \\
    LeanTTS-13M-FM & 16 & 18.898M &
    1.99 & 61.2 & 2.43 & 2.29 & 47.4 & 3.17 \\
    LeanTTS-13M-MF & 4 & 18.898M &
    2.02 & 62.3 & 2.41 & 2.32 & 48.3 & 3.12 \\
    \midrule
    \multicolumn{9}{@{}l}{\textit{Preset/fixed-voice synthesis}} \\
    Matcha-TTS~\cite{mehta2024matcha} & 10 & 32.140M &
    -- & -- & -- & 1.76 & -- & 4.35 \\
    Inflect-Micro-v2~\cite{song2026inflect} & -- & 9.357M &
    -- & -- & -- & 2.77 & -- & 4.41 \\
    LeanTTS-13M-FM & 16 & 18.898M &
    1.56 & 76.7 & 3.11 & 1.88 & 75.3 & 4.34 \\
    LeanTTS-13M-MF & 4 & 18.898M &
    1.57 & 77.2 & 3.03 & 1.93 & 76.6 & 4.33 \\
    \bottomrule
  \end{tabular*}
  \\[2pt]
  \parbox{0.97\textwidth}{\fontsize{9}{9.5}\selectfont
  Params denotes the learned inference path; $\geq$ marks a verified lower
  bound when a complete total is unavailable.}
  \endgroup
\end{table*}